\documentclass[]{spie}  

\usepackage[]{graphicx}
\usepackage[bookmarks = true, pdfpagemode = None, pdfstartview = FitH, colorlinks = true, urlcolor = blue, backref]{hyperref}
\usepackage{cite}
\input{Qcircuit.tex}

\title{Synthesis of Ternary Quantum Logic Circuits by Decomposition}

\author{Faisal Shah Khan\supit{$\dagger$} and Marek M. Perkowski\supit{$*$}
\skiplinehalf
\supit{$\dagger$}   Department of Mathematic and Statistics \\
                    Portland State University, PO Box 751, Portland, Oregon 97207-0751; \\
\supit{$*$}         Department of Electrical and Computer Engineering \\
                    Portland State University, PO Box 751, Portland, Oregon 97207-0751
}

\authorinfo{F.S.K.: E-mail: \url{faisal@pdx.edu}}
\begin{document}
  \maketitle

\begin{abstract}
Recent research in multi-valued logic for quantum computing has shown practical advantages
for scaling up a quantum computer.~\cite{MuthuStroud:04,Sanders:02} Multivalued quantum systems have also been used in the framework of quantum cryptography,~\cite{Pasquinucci:00} and the concept of a qudit cluster state has been proposed by generalizing the qubit cluster state.~\cite{Zhou:03} An evolutionary algorithm based synthesizer for ternary quantum circuits has recently been presented,~\cite{Khan:04} as well as a synthesis method based on matrix factorization~\cite{bullock:04}.In this paper, a recursive synthesis method for ternary quantum circuits based on the Cosine-Sine unitary matrix decomposition is presented.
\end{abstract}




\section{INTRODUCTION}\label{sect:intro}  

A collection of controlled three state quantum systems (qutrits) perturbed by a
classical force can result in the state of one system controlling the evolution of a
second; this is called a quantum circuit. Qutrits replace classical ternary bits as
information units in ternary quantum computing. They are represented as a unit
vector in state space, which is a complex three dimensional vector space, $\mathcal{H}_{3}$.
In the computational basis, the basis vectors (or basis states) of $\mathcal{H}_{3}$ are written in Dirac notation as $\ket{0}$, $\ket{1}$, and $\ket{2}$, where $\ket{0}=(1,0,0)^{T}$ , $\ket{1}=(0,1,0)^{T}$, and $\ket{1}=(0,0,1)^{T}$. An arbitrary vector $\ket{\Psi}$
in $\mathcal{H}_{3}$ can be expressed as a linear combination $\ket{\Psi}=a_{0}\ket{0}+a_{1}\ket{1}+a_{2}\ket{2}$, $a_{0}, a_{1}, a_{2}\in
\textbf{C}$ and $\left|a_{0}\right|^{2}+\left|a_{1}\right|^{2}+\left|a_{2}\right|^{2}=1$.
The real number $\left|a_{i}\right|^{2}$ is the probability that the state vector
$\ket{\Psi}$ will be in  $i$\supit{th} basis state upon measurement.
Note that the basis vectors in the computational basis are ordered by natural numbers.
The state space of ternary quantum system of $n > 1$ qutrits is a composite complex vector
space formed from the algebraic tensor product $\mathcal{H}_{3}^{\otimes{n}}$ of component
state spaces $\mathcal{H}_{3}$. The computational basis for $\mathcal{H}_{3}^{\otimes{n}}$ consists of all possible tensor products of the computational basis vectors of the component spaces; each vector in this basis consists of column vectors with the entry 1 in the $i$-th row and zeros in all others, where $i$ ranges from 1 to $n$.~\cite{Al-Rabadi:04} An arbitrary vector $\ket{\Psi}$ in $\mathcal{H}_{3}^{\otimes{n}}$ can be expressed as linear combination of the basis vectors with scalars $a_{i} \in \textbf{C}$ such that $\sum_{i=0}^{n-1}\left|a_{i}\right|^{2}=1$. The real number  $\left|a_{i}\right|^{2}$ is the probability that the state vector $\ket{\Psi}$ will be in  $i$\supit{th} basis state upon measurement.

The evolution of an $n$ qutrit quantum system occurs via the action of a linear
operator that changes the state vector via multiplication by a $3^{n}\times 3^{n}$
unitary \emph{evolution} matrix. From a computational point of view, the evolution
matrices are quantum logic gates transforming the state vectors in
$\mathcal{H}_{3}^{\otimes n}$. From a quantum logic synthesis point of view, these
gates need to be implemented by a universal set of quantum gates. It is a well-known established
fact that sets of one- and two- qutrit (and in general, \emph{qudit}) gates are
universal~\cite{Brylinski:02,MuthuStroud:04}. Hence, logic synthesis requires that $3^{n}\times 3^{n}$
evolution matrices be efficiently decomposed to the level of one and two qutrit gates. There are several methods from matrix theory, such as QR factorization, that have been utilized
for this purpose in binary quantum logic synthesis. Another method is the Cosine-Sine Decomposition (CSD) of an arbitrary unitary matrix described in section~\ref{sect:CSD}. This method has been recently used by Mottonen~\cite{mottonen:04} et. al, and Shende~\cite{shende:05} et. al for binary logic synthesis. Recently, Bullock et.al have given a synthesis method for multi-valued quantum logic gates using a variation of the QR matrix factorization~\cite{bullock:04}. This paper presents a CSD based method for ternary quantum logic synthesis.


\section{The Cosine-Sine Decomposition (CSD)}\label{sect:CSD}

The Cosine-Sine decomposition has been used recently ~\cite{mottonen:04,shende:05} in the
synthesis of binary quantum gates, which are $2^{n}\times 2^{n}$ unitary matrices for $n$
qubit gates. When used in conjunction with local optimization techniques, the CSD provides
a recursive synthesis method with a lower number of elementary gates compared to other methods~\cite{shende:05}.

\textbf{Cosine-Sine Decomposition}:~\cite{Stewart:77,shende:05} Let the unitary
matrix $\textit{W}\in \textbf{C}^{m\times m}$ be partitioned in $2 \times 2$ block form as
\begin{equation}\label{eqn:CSD matrix}
W=\bordermatrix {  &r      & m-r    \cr
                 r &W_{11} & W_{12} \cr
               m-r &W_{21} & W_{22} \cr}
\end{equation}

\noindent with $2r\leq m$. Then there exist $r \times r$ unitary matrices $U_1,
V_{1}$, $r \times r$ real diagonal matrices $C$ and $S$, and $(m-r) \times (m-r)$ unitary matrices $U_2,V_2$ such that
\begin{equation}\label{eqn:CSD1}
W  = \left(\begin{array}{cc}
  U_{1} & 0 \\ 0 & U_{2}
\end{array}\right)
\left(\begin{array}{ccc}
  C & -S & 0 \\ S & C & 0 \\ 0 & 0 & I_{m-2r}
\end{array}\right)\left(\begin{array}{cc}
  V_{1} & 0 \\  0 & V_{2}
\end{array}\right)
\end{equation}

\noindent The matrices $C$ and $S$ are the so-called cosine-sine matrices and are of the form $C$=diag$(\cos \theta_{1}$, $\cos\theta_{2}$,$\ldots,\cos \theta_{r})$, $S$=diag$(\sin \theta_{1}$, $\sin \theta_{2},\ldots,\sin \theta_{r})$, such that $\sin^{2}\theta_{i}+\cos^{2}\theta_{i}=1$ for $1 \leq i \leq r$.

\textbf{CSD for Binary Quantum Logic Synthesis}: In case of binary quantum logic, all
matrices are even dimensional as powers of two.
Hence, a given $m \times m$ unitary $W$ can be always partitioned into $m/2 \times m/2$
square blocks, giving $m/2 \times m/2$ square matrices $U_{1},U_{2},V_{1},V_{2},C, S$ upon
application of the CSD. The decomposition for this case is given in equation~(\ref{eqn:CSD2}),
\begin{equation}\label{eqn:CSD2}
W  =
\left(\begin{array}{cc}
  U_{1} & 0 \\ 0 & U_{2}
\end{array}\right)
\left(\begin{array}{cc}
  C & -S \\ S & C \\
\end{array}\right)
\left(\begin{array}{cc}
  V_{1} & 0 \\  0 & V_{2}
\end{array}\right)
\end{equation}

\noindent The CSD can be applied to the $2 \times 2$ block diagonal factors that
occur at each iteration until one reaches the qubit level which involves only $2
\times 2$ matrices~\cite{mottonen:04}. At each iteration level the block diagonal
matrices in the decomposition are realized as \emph{quantum
multiplexers}~\cite{shende:05}. A quantum multiplexer is a gate acting on $k$+1
qubits of which one is designated as the control qubit. If the control qubit is the
highest order qubit, the multiplexer matrix is block diagonal. Depending on whether
the control qubit carries $\ket{0}$ or $\ket{1}$, the gate then performs either the
top left block or the bottom right block of the $(k+1) \times (k+1)$ block diagonal
matrix on the remaining $k$ bits. 
\begin{figure}[h]\centerline{
 \Qcircuit @C.4em @R=1em {
 &\qw &\multigate{2}{M} &\qw &\qw & & & &\ & & &\qw &\qw &\qw \qw &\qw &\qw &\qw \qw &\qw &\ctrl{1} &\qw &\qw &\qw &\qw \\
 &\qw &\ghost{M} &\qw &\qw & & &\equiv & & & &\qw &\qw&{/} &\qw &\qw &\gate{\textbf{F}} &\qw &\multigate{1}{M} &\qw &\qw &\qw \\
 &\qw &\ghost{M} &\qw &\qw & & & & & & &\qw &\qw &{/} &\qw &\qw \qwx &\gate{\textbf{G}} &\qw &\ghost{M} &\qw &\qw &\qw \\}}
 \caption{A 3-qubit quantum multiplexer. The / represents two wires, one for each lower qubit. Depending on the value of the controlling qubit, \textbf{F} or \textbf{G} is applied to the lower two qubits.}
\end{figure}
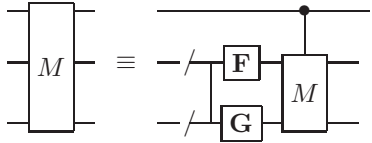

For instance, a quantum multiplexer matrix for 3 qubits will be a $2 \times 2$ block diagonal matrix with each block matrix of size $4 \times 4$ given in equation~(\ref{eqn:3qubits mux}). The value of the first qubit is $\ket{0}$ in the location of the block matrix \textbf{F} and $\ket{1}$ in the location of the block matrix \textbf{G}. Therefore, depending on whether the control bit carries $\ket{0}$ or $\ket{1}$, the gate then performs either \textbf{F} or \textbf{G} on the remaining $2$ qubits respectively. 

\begin{equation}\label{eqn:3qubits mux}
\left(\begin{array}{cc} \textbf{F} & 0  \\ 0 & \textbf{G}
\end{array}\right)
\end{equation}

The cosine-sine matrices in the CSD are realized as \emph{uniformly k-controlled}
$R_{y}$ rotations~\cite{shende:05,mottonen:04}. Such gates operate on $k + 1$ qubits, of which the lower $k$ are controls and the top one is the target. A different $R_y$ is applied to the target for each control bit-string. The circuit for a uniformly 2-controlled $R_{y}$ rotation is given in figure 2. 
\begin{figure}[h]\centerline{
 \Qcircuit @C1em @R=1.5em {
&\gate{R_{y}} &\qw & & & & &\qw &\qw &\gate{R_{y}^{\theta_{0}}} &\qw &\gate{R_{y}^{\theta_{1}}} &\qw &\gate{R_{y}^{\theta_{2}}} &\qw &\gate{R_{y}^{\theta_{3}}} &\qw &\qw &\qw \\
&\ctrl{-1} &\qw & &\equiv & & &\qw &\qw &\ctrlo{-1} &\qw &\ctrl{-1} &\qw &\ctrlo{-1} &\qw &\ctrl{-1} &\qw &\qw &\qw\\
 &\ctrl{-1} &\qw & & & & &\qw &\qw &\ctrlo{-1} &\qw &\ctrlo{-1} &\qw &\ctrl{-1} &\qw &\ctrl{-1} &\qw &\qw &\qw\\}}
\caption{A uniformly 2-controlled $R_{y}$ rotation: the lower two bits are the control bits, and the top bit is the target bit. In general, it requires $2^{k}$ one qubit controlled gates to implement a uniformly $k$-controlled rotation. The open circles represent the value $\ket{0}$ and the closed circles the value $\ket{1}$.}
\end{figure}
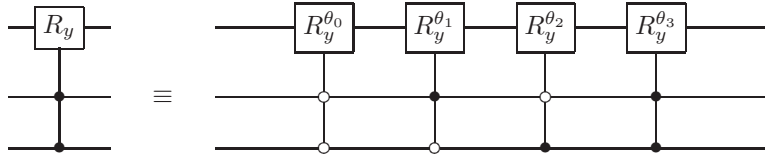

For three qubits $\ket{a}, \ket{b}$, and $\ket{c}$, the matrix formulation of a uniformly 2-controlled $R_{y}$ rotation gate is given in equation~(\ref{eqn:three qubit cosinesine}).

\begin{equation}\label{eqn:three qubit cosinesine}
\left(\begin{array}{cc} \cos\theta_{i} & -\sin\theta_{i} \\ \sin\theta_{i} &\cos\theta_{i} \\
\end{array}\right)
\left(\begin{array}{c} a_{0} \\ a_{1} \\
\end{array}\right)
\otimes\left(\begin{array}{c} b_{0} \\ b_{1} \\
\end{array}\right)
\otimes\left(\begin{array}{c} c_{0} \\ c_{1} \\
\end{array}\right)
\end{equation}

\noindent As $\ket{b}$ and $\ket{c}$ in equation~(\ref{eqn:three qubit cosinesine}) take on the values $\ket{0}$ and $\ket{1}$ in four possible combinations, $\theta_{i}$ takes on the values from the set $\left\{\theta_{0},\theta_{1},\theta_{2}, \theta_{3}\right\}$, resulting in different $R_{y}$ gates being applied to the top most qubit. Each $\theta_{i}$ is an arbitrary angle.

\section{CSD for Ternary Quantum Logic Synthesis}\label{sec:Ternary CSD}
For ternary quantum logic synthesis, an $n$-qutrit gate will be a unitary matrix $W$ of size
$3^{n} \times 3^{n}$. Partition $W$ as in (1) with $m=3^{n}$ and $r=3^{n-1}$, so
that $m-r = 3^{n}-3^{n-1}=3^{n-1}(3-1)=3^{n-1}\cdot2$. After the application of the
CSD, $W$ will take the form in equation~(\ref{eqn:CSD1}). The matrix blocks $U_{2},V_{2}$ will be of size $3^{n-1}\cdot2 \times 3^{n-1}\cdot2 $; hence an application of the CSD only on these two blocks will decompose each block into the form in equation~(\ref{eqn:CSD2}). After these two application of the CSD and some matrix factoring, $W$ will take the form

\begin{equation}\label{eqn:Ternary Decomposed CSD}
W=
\Sigma
\left(\begin{array}{ccc}
  C & -S & 0 \\ S & C & 0 \\  0 & 0 & I
\end{array}\right)
\Gamma
\end{equation}
with
\begin{equation}\label{eqn:Sigma}
\Sigma =
\left(\begin{array}{ccc}
  X_{1} & 0 & 0 \\ 0 & X_{2} & 0 \\  0 & 0 & X_{3}
\end{array}\right)
\left(\begin{array}{ccc}
  I & 0 & 0 \\ 0 & C_{1} & -S_{1} \\  0 & S_{1} & C_{1}
\end{array}\right)
\left(\begin{array}{ccc}
  I & 0 & 0 \\ 0 & Z_{1} & 0 \\  0 & 0 & Z_{2}
\end{array}\right)
\end{equation}
and
\begin{equation}\label{eqn:Gamma}
\Gamma=
\left(\begin{array}{ccc}
  Y_{1} & 0 & 0 \\ 0 & Y_{2} & 0 \\ 0 & 0 & Y_{3}
\end{array}\right)
\left(\begin{array}{ccc}
  I & 0 & 0 \\ 0 & C_{2} & -S_{2} \\ 0 & S_{2} & C_{2} \\
\end{array}\right)
\left(\begin{array}{ccc}
  I & 0 & 0 \\ 0 & W_{1} & 0 \\  0 & 0 & W_{2}
\end{array}\right)
\end{equation}

Each block matrix in the decomposition given in equations~(\ref{eqn:Ternary Decomposed CSD}) -~(\ref{eqn:Gamma}) above is of size $3^{n-1}\times
3^{n-1}$. We realize each block diagonal matrix as a ternary quantum multiplexer
acting on $n$ qutrits of which the highest order qutrit is designated as the control
qutrit. Depending on which of the  values $\ket{0}$, $\ket{1}$, or $\ket{2}$ the control qutrit carries, the gate then performs either the top left block, the middle block, or the bottom right block respectively on the remaining $n-1$ qutrits. The cosine-sine matrices with identity in top-left/bottom-right block corner are realized as uniformly $(n-1)$-controlled $R_{x}/R_{z}$ rotations. These matrices can be realized as $R_{x}$ or $R_{z}$ rotation~\cite{Fuji:03} matrices in $\textbf{R}^{3}$ applied to the top most qutrit, controlled by the lower qutrits as they range over $\left\{\ket{0}, \ket{1}, \ket{2}\right\}$. Each configuration of the lower qutrits leads to a different $R_{x}$ or $R_{z}$ gate.

\textbf{EXAMPLE} Consider two qutrits being acted upon by an arbitrary gate $Q$. The CSD synthesis of $Q$ is given in figure 3.
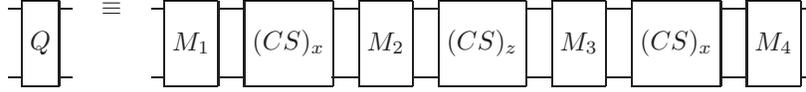
\begin{figure}[h]\centerline{
\Qcircuit @C.5em @R=2em {& &\multigate{1}{Q} &\qw & & &\equiv & & & &\multigate{1}{M_{1}} &\qw &\multigate{1}{(CS)_{x}} &\qw &\multigate{1}{M_{2}} &\qw  &\multigate{1}{(CS)_{z}} &\qw &\multigate{1}{M_{3}} &\qw &\multigate{1}{(CS)_{x}} &\qw &\multigate{1}{M_{4}} &\qw \\
& &\ghost{Q} &\qw & & & & & & &\ghost{M_{1}} &\qw &\ghost{(CS)_{1}} &\qw &\ghost{M_{2}} &\qw &\ghost{(CS)_{2}} &\qw &\ghost{M_{3}} &\qw &\ghost{(CS)_{3}} &\qw  &\ghost{M_{4}} &\qw \\}}\caption{The decomposition of an arbitrary 2-qutrit gate $Q$ using the CSD. Each $M_{i}$ is a ternary quantum multiplexer. The gates $(CS)_{x}$ and $(CS)_{z}$ are uniformly 1-controlled rotations}
\end{figure}

For $1 \leq i \leq 4$, each $M_{i}$ gate in figure 3 is a quantum multiplexer controlled by the top qutrit and can be decomposed to the level of elementary gates as shown in figure 4.

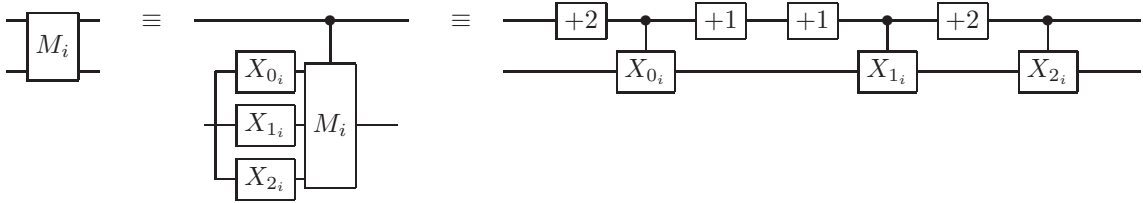
\begin{figure}[h]\centerline{
 \Qcircuit @C.4em @R=.5em {
&\qw &\multigate{1}{M_{i}} &\qw &\qw & & & & &\equiv & & & & &\qw &\qw &\qw &\qw \qw &\ctrl{1} &\qw &\qw &\qw &\qw &\qw & & & & &\equiv & & & & &\qw &\qw &\qw &\qw &\gate{+2} \qw &\ctrl{1} &\qw &\gate{+1} &\qw &\qw &\qw &\gate{+1} &\qw &\ctrl{1} &\qw &\gate{+2} &\qw &\qw &\ctrl{1} &\qw &\qw &\qw &\qw &\qw &\qw\\
&\qw &\ghost{M_{i}} &\qw &\qw & & & & & & & & & & & &\qw &\gate{X_{0_{i}}} &\multigate{2}{M_{i}} & & & & & & & & & & & & & & &\qw &\qw &\qw &\qw &\qw &\gate{X_{0_{i}}}  &\qw &\qw &\qw &\qw &\qw &\qw &\qw &\gate{X_{1_{i}}}  &\qw &\qw &\qw &\qw &\gate{X_{2_{i}}} &\qw &\qw &\qw &\qw &\qw &\qw\\
& & & & & & & & & & &&  & & &\qw\qwx &\qw &\gate{X_{1_{i}}} &\ghost{M_{i}} &\qw &\qw &\qw &\qw \\
& & & & &  & & & & & & & & & &\qwx &\qw &\gate{X_{2_{i}}}&\ghost{M_{i}} & & & & & \\}}\caption{Quantum Ternary Multiplexer for second qutrit and its realization in terms of Muthukrishan-Stroud gates. The gates labled +1 and +2 are bit shifts increasing the value of the bit by 1 and 2 mod 3 respectively. Depending on the value of the top control qutrit $a$, one of $X_{a_{i}}$ is applied to the second qutrit, for $a\in\left\{0,1,2\right\}$.}
\end{figure}

For two qutrits, the matrix for a ternary quantum multiplexer will be a $3\times3$ block diagonal matrix given in equation~(\ref{eqn:two qutrit MUX matrix}). The value of the first qutrit is $\ket{0}$ in the location of the block matrix \textbf{F}, $\ket{1}$ in the location of the block matrix \textbf{G}, and $\ket{2}$ in the location of the block matrix \textbf{H}. Therefore, depending on whether the control bit carries $\ket{0}$, $\ket{1}$, or $\ket{2}$, the gate then performs either \textbf{F}, \textbf{G}, or \textbf{H} on the remaining qutrit respectively. All blocks in the matrix in equation~(\ref{eqn:two qutrit MUX matrix})are of size $3 \times 3$.

\begin{equation}\label{eqn:two qutrit MUX matrix}
\left(\begin{array}{ccc} \textbf{F} & 0 & 0 \\ 0 & \textbf{G} & 0 \\ 0 & 0 & \textbf{H}
\end{array}\right)
\end{equation}

The gates $(CS)_{x}$ and $(CS)_{z}$ in figure 3 are uniformly 1-controlled $R_{x}$ and $R_{z}$ rotations respectively. In either case, the top qutrit is controlled by the lower one, as shown in figure 5. The gate $(CS)_{z}$ corresponds to the middle matrix in equation~(\ref{eqn:Ternary Decomposed CSD}). The matrix formulation of this gate as a uniformly 1-controlled $R_{z}$ rotation is given in equation~(\ref{eqn:two qutrit cosinesine}).

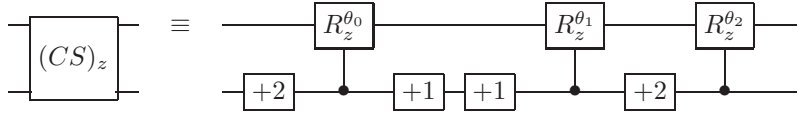
\begin{figure}[h] \centerline{
 \Qcircuit @C.4em @R=1em{
&\qw &\multigate{1}{(CS)_{z}} &\qw &\qw & & & & \equiv & & & & & \qw &\qw &\qw &\gate{R_{z}^{\theta_{0}}} \qw &\qw &\qw &\qw &\qw &\qw &\qw &\gate{R_{z}^{\theta_{1}}} &\qw &\qw &\qw &\gate{R_{z}^{\theta_{2}}} &\qw &\qw &\qw &\qw\\
&\qw &\ghost{(CS)_{i}} &\qw &\qw & & & & & & & & &\qw &\gate{+2}
&\qw &\ctrl{-1} &\qw &\gate{+1} &\qw &\gate{+1} &\qw &\qw &\ctrl{-1} &\qw &\gate{+2} &\qw &\ctrl{-1} &\qw &\qw &\qw &\qw\\}}\caption{Uniformly
1-controlled rotations for 2-qutrits, realized as multiplexers via Muthukrishan-Stroud gates. The gates labeled +1 and +2 are bit shifts modulo 3.}
\end{figure}

\begin{equation}\label{eqn:two qutrit cosinesine}
\left(\begin{array}{ccc} \cos\theta_{i} & -\sin\theta_{i} & 0\\ \sin\theta_{i} &\cos\theta_{i} & 0 \\ 0 & 0 & 1
\end{array}\right)
\left(\begin{array}{c} a_{0} \\ a_{1} \\ a_{2}
\end{array}\right)
\otimes\left(\begin{array}{c} b_{0} \\ b_{1} \\ b_{2}
\end{array}\right)
\end{equation}

\noindent Depending on the three possible binary configurations of $\ket{b}$, $\theta_{i}$ takes on the values from the set $\left\{\theta_{0},\theta_{1},\theta_{2}\right\}$, resulting in different $R_{z}$ gates being applied to the top qutrit. If the lower qutrit is $\ket{0}$, equation~(\ref{eqn:two qutrit cosinesine}) reduces to

\begin{equation}\label{eqn:two qutrit cosinesine1}
\left(\begin{array}{ccc} \cos\theta_{1} & -\sin\theta_{1} & 0 \\ \sin\theta_{1} &\cos\theta_{1} & 0 \\
0 & 0 & 1\\
\end{array}\right)
\left(\begin{array}{c} a_{1} \\ a_{2} \\ a_{3}\\
\end{array}\right)
\otimes\left(\begin{array}{c} 1 \\ 0 \\ 0
\\
\end{array}\right)
\end{equation}

If the lower qutrit is $\ket{1}$ or $\ket{2}$, equation~(\ref{eqn:two qutrit cosinesine})
reduces to equations~(\ref{eqn:two qutrit cosinesine2}) and~(\ref{eqn:two qutrit
cosinesine3}) respectively.

\begin{equation}\label{eqn:two qutrit cosinesine2}
\left(\begin{array}{ccc} \cos\theta_{2} & -\sin\theta_{2} & 0 \\ \sin\theta_{2} &\cos\theta_{2} & 0 \\
0 & 0 & 1\\
\end{array}\right)
\left(\begin{array}{c} a_{1} \\ a_{2} \\ a_{3}\\
\end{array}\right)
\otimes\left(\begin{array}{c} 0 \\ 1 \\ 0 \\
\end{array}\right)
\end{equation}

\begin{equation}\label{eqn:two qutrit cosinesine3}
\left(\begin{array}{ccc} \cos\theta_{3} & -\sin\theta_{3} & 0 \\ \sin\theta_{3} &\cos\theta_{3} & 0 \\
0 & 0 & 1\\
\end{array}\right)
\left(\begin{array}{c} a_{1} \\ a_{2} \\ a_{3}\\
\end{array}\right)
\otimes\left(\begin{array}{c} 0 \\ 0 \\ 1
\\
\end{array}\right)
\end{equation}

Hence, a 2-qutrit quantum gate can be synthesized via four $1$ qutrit quantum
multiplexers and three 1-qutrit uniformly controlled rotations on the first qutrit. In general, an $n$-qutrit quantum gate can be synthesized via four $n-1$ qutrit quantum
multiplexers and three uniformly $n-1$ controlled rotations on the top qutrit. 
\section{Conclusions and Future Work}\label{sec:Conclusions}
We give a recursive procedure for ternary quantum logic synthesis by
realizing $n$ qutrit logic gates as $3^{n}\times 3^{n}$ unitary matrices
and applying the Cosine-Sine Decomposition. We conclude that this method
can synthesize a $n$ qutrit gate with four multiplexers acting on $n-1$ qutrits
and three uniformly $n-1$-controlled rotations. A two qutrit example is given. It is our future goal to do a gate count by investigating local optimizations at each level of recursion. We also intend to write a CAD tool for this decomposition and get a gate count for a higher number of qutrits, and extend the decomposition to odd radix multi-valued quantum logic synthesis.
\section{Acknowledgments}\label{sec:Acknowledgment}

F.~S.~Khan is grateful to Jacob Biamonte for discussions, advice,
and help in the layout of this paper. The Quantum Circuit diagrams
were all drawn in \LaTeX \ using Q-circuit available at
\href{http://info.phys.unm.edu/Qcircuit/}{http://info.phys.unm.edu/Qcircuit/}.


\end{document}